\documentclass[aps,prc,groupedaddress,showpacs,manuscript]{revtex4}

\usepackage{amsmath}
\usepackage{graphicx}

\begin{document}

\title{System-size dependence of the pion freeze-out volume as a potential signature for the phase transition to a Quark Gluon Plasma}

\author {Qingfeng Li$\, ^{1}$\footnote{E-mail address:
liqf@hutc.zj.cn}, Caiwan Shen$\, ^{1}$, and Marcus Bleicher$\,
^{2,3}$}
\address{
1) School of Science, Huzhou Teachers College, Huzhou 313000,
People's Republic of China \\
2) Frankfurt Institute for Advanced Studies (FIAS), Johann Wolfgang Goethe-Universit\"{a}t, Max-von-Laue-Str.\ 1, D-60438 Frankfurt am Main, Germany\\
3) Institut f\"{u}r Theoretische Physik, Johann Wolfgang Goethe-Universit\"{a}t, Max-von-Laue-Str.\ 1, D-60438 Frankfurt am Main, Germany\\
\\
 }


\begin{abstract}
Hanburry-Brown-Twiss (HBT) correlation functions and radii of
negatively charged pions from C+C, Si+Si, Cu+Cu, and In+In at lower
RHIC/SPS energies are calculated with the UrQMD transport model and
the CRAB analyzing program. We find a minimum in the excitation
function of the pion freeze-out volume at low transverse momenta and
around $E_{lab}\sim 20-30A$GeV which can be related to the
transition from hadronic to string matter (which might be
interpreted as a pre-cursor of the QGP). The existence of the
minimum is explained by the competition of two mechanisms of the
particle production, resonance decays and string
formation/fragmentation.
\end{abstract}


 \maketitle

Nowadays one expects that a (phase) transition from a hadron gas
(HG) to the quark-gluon-plasma (QGP) is encountered in relativistic
heavy-ion collisions (HICs). Since more than a decade it has been
suggested that the onset of deconfinement happens in HICs at
SPS/FAIR energies, an energy regime which is now also at the focus
of the current RHIC low energy scan program. According to current
lattice QCD calculations one expects - in order of increasing
baryo-chemical potential - a cross over transition, a critical
endpoint with a second order transition and for high baryon
densities a first order phase transition. In fact, the combination
of theoretically predicted signals and corresponding data, has led
to the emergence of a complete phase diagram of strongly interacting
matter
\cite{Matsui:1986dk,Soff:1999et,Dumitru:2001xa,Gazdzicki:2006fy,Heinz:2006ur,Grebieszkow:2009th,Hohne:2009yw,BraunMunzinger:2009zz,Andronic:2009gj}.
Unfortunately, uncertainties still exist and the present data does
not seem to allow for firm conclusion, especially when it comes to
the location and observation of the critical point
\cite{Schuchmann07,Grebieszkow:2009th,Hohne:2009yw,Konchakovski:2010fh}.
To reduce these uncertainties and to draw final conclusions about
the location of the critical endpoint and the boundary of the first
order transition, a more careful beam-energy scan of these signals
for the onset of deconfinement is currently undertaken and planned
for the future. Current experimental efforts are centered at several
international laboratories such as GSI at Darmstadt, CERN at Geneva,
and RHIC at BNL and will go on with the SIS-100(300)
\cite{Hohne:2009yw}, the NA61/SHINE
\cite{Gazdzicki:2006fy,Gazdzicki:2008pu,Grebieszkow:2009th}, and the
critRHIC \cite{Stephans:2006tg,Stephans:2008zz} program.

In the present paper we go back to a long standing idea to explore
the regions of homogeneity by particle correlations, namely the
Hanbury-Brown-Twiss (HBT) technique
\cite{HBT54,Goldhaber60,Bauer:1993wq}. The HBT technique allows to
extract information on the spatio-temporal evolution of the source
of various two-particle species. A plethora of data on the whole
beam-energy scan from AGS \cite{Lisa:2000hw,Ahle:2002mi}, over
CERN-SPS \cite{Kniege:2004pt,Kniege:2006in,Adamova:2002wi} up to
RHIC energies
\cite{Back:2004ug,Adler:2001zd,Adcox:2002uc,Adler:2004rq,Adams:2004yc}
exists, however, mainly for the pion source from heavy systems. In
Ref.\ \cite{Adamova:2002ff} a systematic analysis of the pion
freeze-out data at beam energies from AGS to RHIC is laid out for
the first time and they found a steep decrease of the pion
freeze-out volume at AGS energies and an increase throughout the SPS
energy regime towards RHIC. It indicates the existence of a minimum
between AGS and SPS energies and it is claimed that this is due to
the transition from nucleon to pion dominated freeze-out between AGS
and SPS energies. However, this behaviour can not be explained by
transport model calculations \cite{Lisa:2005dd,Li:2007im} within a
cascade mode. Furthermore, the systematic and the statistic errors
of the data are still large (for the apparently rather large
systematic uncertainties, we refer the reader to the differences
between the NA49 and the CERES data \cite{Alt:2007uj,Schuchmann07})
and it is therefore difficult to draw firm conclusion based on the
currently existing data \cite{Alt:2007uj,Hohne:2009yw}.

When neglecting the strong space-time correlation, it is known that
the ratio between the HBT radii in outward and sideward directions
($R_{O}, R_{S}$) is related to the emission duration of the source.
Thus, a ratio larger than unity is expected when the system crosses
the first-order phase transition \cite{Rischke:1996em} and remains
for a long time in the mixed phase. However, an almost unity value
has been shown by experiments throughout the beam energies from AGS
to RHIC \cite{Lisa:2005dd}. This seems to imply a strong
space-momentum correlation \cite{Li:2007yd} which has to be taken
into account in the interpretation of the experimental data. In
recent years both the non-monotonous energy dependence of the
freeze-out volume and the small $R_{O}/R_{S}$-ratio present in the
AGS, SPS, and RHIC energy regions have been explained fairly well by
both analyzing non-Gaussian effects and by considering a stronger
early pressure which comes from the contribution of mean-field
potentials for both formed and preformed hadrons
\cite{Li:2007yd,Li:2008ge,Li:2008bk,Pratt:2009hu}. Recently, it was
pointed out \cite{Lizhu:2010wy} that a sufficient spatial size of
the fire ball is essential to locate the critical point, the
crossover and the first-order phase transition in the phase diagram.

On the transport model side, benchmark test have been performed,
which provide a solid basis for the present and future
investigations. The results obtained previously are in line with the
experimental facts that the extracted HBT radii are always rather
small and change smoothly for the large span of explored beam
energies. I.e. no unexpected observations about the freeze-out
volume of two particle correlations is found at low SPS energies
where the energy threshold for the onset of deconfinement might be
reached. This implies that the expected sensitivity of the pion
freeze-out volume to the possible phase transition is relatively
weak and needs to be analyzed carefully. On the experimental side a
detailed exploration of different collision systems is currently
underway with the NA61 experiment taking data at beam energies from
10A GeV to 158A GeV for systems from p+p to Au+Au
\cite{Gazdzicki:2006fy,Gazdzicki:2008pu}. Finally, we suggest to
conduct a careful transverse-momentum analysis for the excitation
function of the pion freeze-out volume in order to find out any
although weak but unusual phenomenon for the phase transition.

In this work, we provide a baseline calculation for the system-size and the
transverse-momentum dependence of the excitation function of the HBT pion freeze-out volume
within a relativistic transport model. In line with our previous findings, we explore two
scenarios, the standard cascade calculations and a modified version of the model which
includes mean field potentials of both formed and preformed hadrons to increase the pressure
in the early stage of the reaction. As a tool we employ a modified version of the UrQMD model
which - apart from other changes - includes a repulsive interaction for the early stage of
the reaction to mimic the explosive expansion of the source encountered at the highest
energies (for details of the implementation, the reader is referred to
\cite{Li:2007yd,Li:2008ge,Li:2008bk}).

\begin{figure}
\includegraphics[angle=0,width=0.8\textwidth]{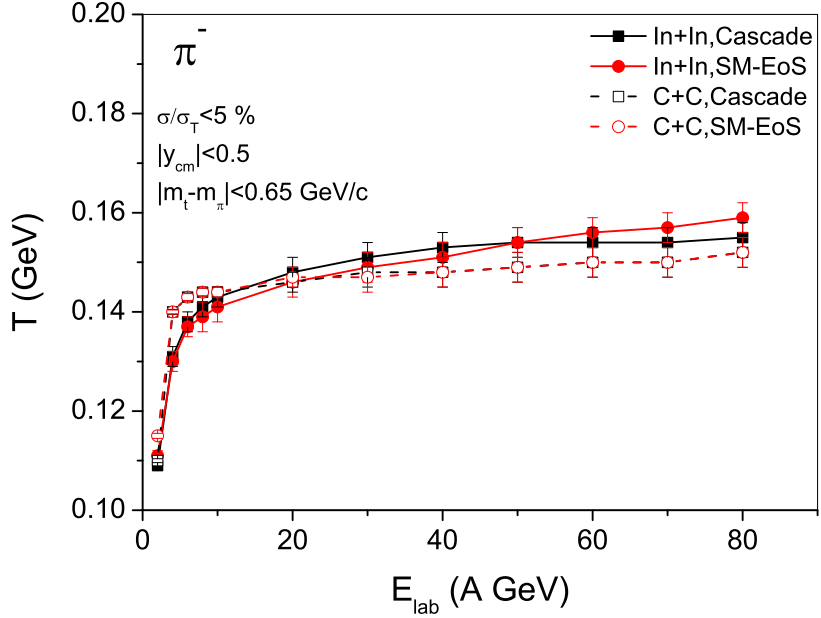}
\caption{Excitation function of the extracted temperature parameter $T$ of negatively charged
pions at mid-rapidity from central C+C and In+In reactions at AGS and SPS energies. The
comparison of calculations between with and without mean-field potentials is shown for each
colliding system.} \label{fig1}
\end{figure}

For the present analysis we simulate central collisions ($\sigma/\sigma_{Total}<5\%$,
$\sigma$ being the cross section) for four mass-symmetric reactions, C+C, Si+Si, Cu+Cu, and
In+In, with beam energies from 2A GeV to 80A GeV (for detailed energy points, please see the
figures below).  Firstly we extract the inverse slope parameter $T$ (``apparent temperature''
or ``temperature'') from the transverse mass ($m_t=(m^2+p_t^2)^{1/2}$) spectra of negatively
charged pions at mid-rapidity ($|y_{cm}|<0.5$, where
$y_{cm}=\frac{1}{2}\log{\frac{E+p_{\parallel}}{E-p_{\parallel}}}$, $E$ and $p_\parallel$ are
the energy and longitudinal momentum of the pion meson in the center-of-mass system)
according to the expression
\begin{equation}
\frac{1}{m_t}\frac{dN^2}{dm_tdy_{cm}}= f(y_{cm})
\exp(-\frac{m_t}{T}) \label{eq1}
\end{equation}
where $f(y_{cm})=const$. Fig.\ \ref{fig1} shows the excitation function of the extracted $T$
values at AGS and SPS energies (up to 80A GeV). Calculations with (``SM-EoS'') and without
(``Cascade'') mean-field potentials of both formed and preformed hadrons from C+C and In+In
reactions are compared with each other. In order to extract the temperature parameter $T$ a
transverse mass upper limit is chosen to avoid potential problems with deviations from a
single exponential spectrum. Here we use the range $|m_t-m_\pi|<0.65$GeV$/c$ for the fitting
process, in line with the transverse momentum range for the HBT analysis later on.

As shown in previous UrQMD calculations and the experimental results
\cite{Bratkovskaya:2004kv}, the excitation function of the extracted
temperature from pion spectra is weakly dependent on the size of the
system, which is shown in Fig.\ \ref{fig1}. At AGS a rapid increase
of the $T$ from about $0.11$ GeV to $0.14$ GeV with increasing beam
energies is seen. With the further increase of beam energies, the
$T$ becomes flat and shows no obvious beam-energy dependence
especially for the light system and in cascade calculations. From
the results shown in Fig.\ \ref{fig1} one also finds that the
mean-field potential contribution to the modification of $T$ is also
of minor importance and disappears completely for the light C+C
system. Note that the mean-field potential for pre-formed hadrons
(string fragments) is also considered in the present calculations.
Therefore, the weak dependence of the apparent temperature on
different equations of state in such a large beam-energy region
implies that most of the previously predicted signatures to the
confinement-deconfinement phase transition would not be that bright
in the real dynamic transport process
\cite{Haussler:2007un,Petersen:2010md}.

Let us now turn to the correlation's study. The program Correlation
After Burner (CRAB) (version 3.0$\beta$)
\cite{Pratt:1994uf,Pratthome} is used to analyze the two-particle
interferometry. The correlator $C$ of two particles is decomposed in
Pratt's (so-called longitudinal co-moving system LCMS or
``Out-Side-Long'') three-dimensional convention (Pratt-radii). The
three-dimensional correlation function is fit with the standard
Gaussian form (using ROOT \cite{roothomepage} and minimizing the
$\chi$-squared)
\begin{equation}
C(q_L,q_O,q_S)=1+\lambda
e^{-R_L^2q_L^2-R_O^2q_O^2-R_S^2q_S^2-2R_{OL}^2q_Oq_L}. \label{fit1}
\end{equation}
In Eq.~(\ref{fit1}), $\lambda$ is usually referred to as an incoherence factor. However, it
might also be affected by many other factors, such as the contaminations from long-lived
resonances or the details of the Coulomb modification in final state interactions (FSI). Here
we regard it as a free parameter and do not assign a specific physical meaning to it.  $R_L$,
$R_O$, and $R_S$ are the Pratt-radii in longitudinal, outward, and sideward directions, while
the cross-term $R_{OL}$ plays a role at large rapidities. $q_i$ is the pair relative momentum
$\mathbf{q}$ ($\mathbf{q}=\mathbf{p}_1-\mathbf{p}_2$) in the $i$ direction. Furthermore, the
correlator is also $k$-dependent (the transverse component is preferred under a rapidity
cut), where $k=(p_1+p_2)/2$ is the average momentum of the two particles.

\begin{figure}
\includegraphics[angle=0,width=0.8\textwidth]{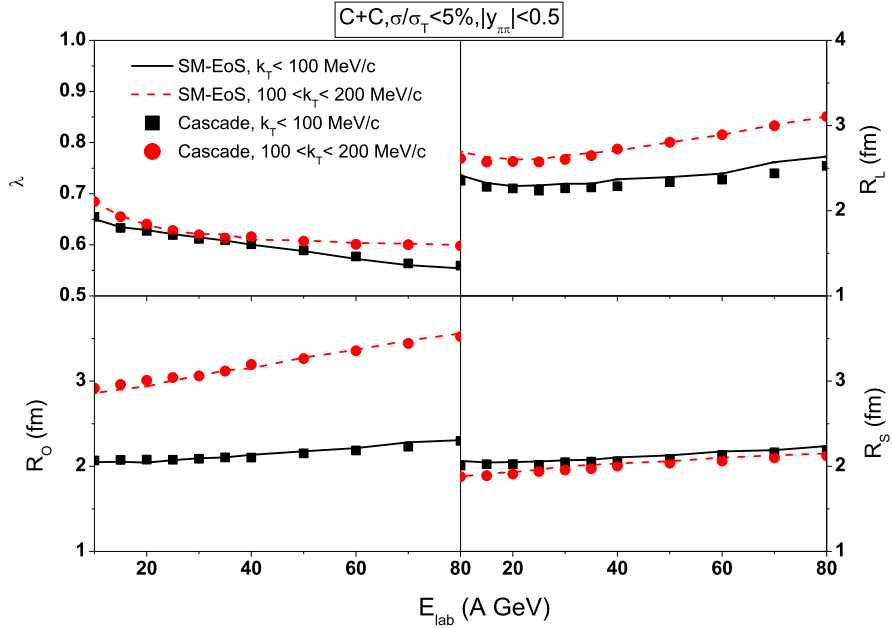}
\caption{Excitation functions of the HBT parameters $\lambda$ (upper-left), $R_L$
(upper-right), $R_O$ (bottom-left), and $R_S$ (bottom-right) of $\pi^--\pi^-$ pairs from
central C+C collisions at two $k_T$ bins $0-100$ MeV$/c$ and $100-200$ MeV$/c$. A pion-pair
rapidity cut $|Y_{\pi\pi}|<0.5$ is employed. Calculations in the cascade mode are depicted by
symbols while calculations with SM-EoS are shown by the different lines.} \label{fig2}
\end{figure}

In following calculations, $10^9$ negatively charged pion pairs are calculated at
mid-rapidity with the two correlated particles at $|Y_{\pi\pi}|<0.5$ (where
$Y_{\pi\pi}=\frac{1}{2}{\rm log}(\frac{E_1+E_2+p_{\parallel 1}+p_{\parallel
2}}{E_1+E_2-p_{\parallel 1}-p_{\parallel 2}})$ is the pair rapidity with pion energies $E_1$
and $E_2$ and longitudinal momenta $p_{\parallel 1}$ and $p_{\parallel 2}$ in the center of
mass system) in each CRAB analyzing run. Fig.\ \ref{fig2} depicts excitation functions of the
HBT parameters $\lambda$ (upper-left), $R_L$ (upper-right), $R_O$ (bottom-left), and $R_S$
(bottom-right) of $\pi^--\pi^-$ pairs from central C+C collisions at two $k_T$
($=(p_{1T}+p_{2T})/2$) bins $0-100$ MeV$/c$ and $100-200$ MeV$/c$. Calculations with cascade
mode are shown with scattered symbols while calculations with SM-EoS are shown by different
lines.

\begin{figure}
\includegraphics[angle=0,width=0.8\textwidth]{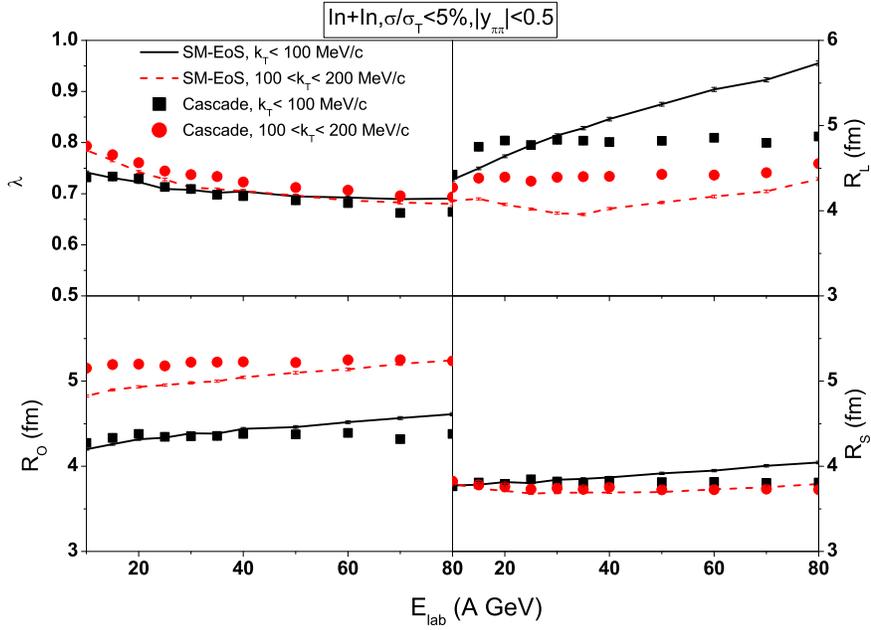}
\caption{Excitation functions of the HBT parameters $\lambda$ (upper-left), $R_L$
(upper-right), $R_O$ (bottom-left), and $R_S$ (bottom-right) of $\pi^--\pi^-$ pairs from
central In+In collisions at two $k_T$ bins $0-100$ MeV$/c$ and $100-200$ MeV$/c$. A pion-pair
rapidity cut $|Y_{\pi\pi}|<0.5$ is employed. Calculations in the cascade mode are depicted by
symbols while calculations with SM-EoS are shown by the different lines.} \label{fig3}
\end{figure}

Similarly, Fig.\ \ref{fig3} gives the results of In+In collisions. A
beam-energy range $10$A-$80$A GeV is selected and the number of
calculated energy points is large enough in order to give clearer
information about the energy dependence of these HBT parameters. The
$\lambda$ parameter is less than unity for both systems at AGS and
SPS energies and decreases with increasing beam energy. This
behaviour is related to the increasing importance of long lived
resonances and substantial rescattering. It was shown that the
excitation function of the experimental $\lambda$ values from Au+Au
collisions at AGS energies can be reproduced well by the RQMD model
\cite{Lisa:2000hw}. Although the effect of the mean-field potential
on $\lambda$ is weak in both reactions at all beam energies as shown
in \cite{Lisa:2000hw}, it indeed becomes visible in the heavier
In+In system especially at high SPS energies. We further find that
at the low $k_T$ bin the $\lambda$ value is driven up slightly when
potentials are considered and vice versa for the high $k_T$ bin.
This is easy to understand because the attractive potential leads to
more rescattering process for pions with small momenta, which
implies a larger incoherence. While the repulsive part of the
potential leads to less rescattering of the resonances before
freeze-out, which implies thus a larger coherence. If we turn to
look into the beam-energy dependence of the HBT radii, surprisingly,
we find some non-monotonous behaviour at low SPS energies,
especially in the longitudinal direction. In C+C collisions and
within the $k_T$ bin $0-100$ MeV$/c$, a minimum $R_L$ is seen at
$\sim 25A$ GeV for calculations both with and without potentials.
While the appearance of the minimum value of $R_L$ from In+In
collisions depends heavily on both the consideration of the
potential and the $k_T$ interval selected - the minimum appears at
$\sim 35A$ GeV only from the calculation with potentials and at the
larger $k_T$ bin $100-200$ MeV$/c$. Furthermore, a stronger
beam-energy dependence of the HBT radii with mean-field potentials
is also seen especially in the longitudinal direction of pion pairs
from In+In collisions which is certainly due to the density
dependence of the Skyrme-like terms in SM-EoS \cite{Li:2007yd}.
Therefore, it is seen clearly that for light system, the minimum
point appears at both $k_T$ bins, while for heavy system, the
appearance of the minimum depends on the consideration of both the
certain $k_T$ bin and the mean-field potentials. This is interesting
phenomenon which can be addressed by the system size and energy scan
of the NA61 experiment.

\begin{figure}
\includegraphics[angle=0,width=0.8\textwidth]{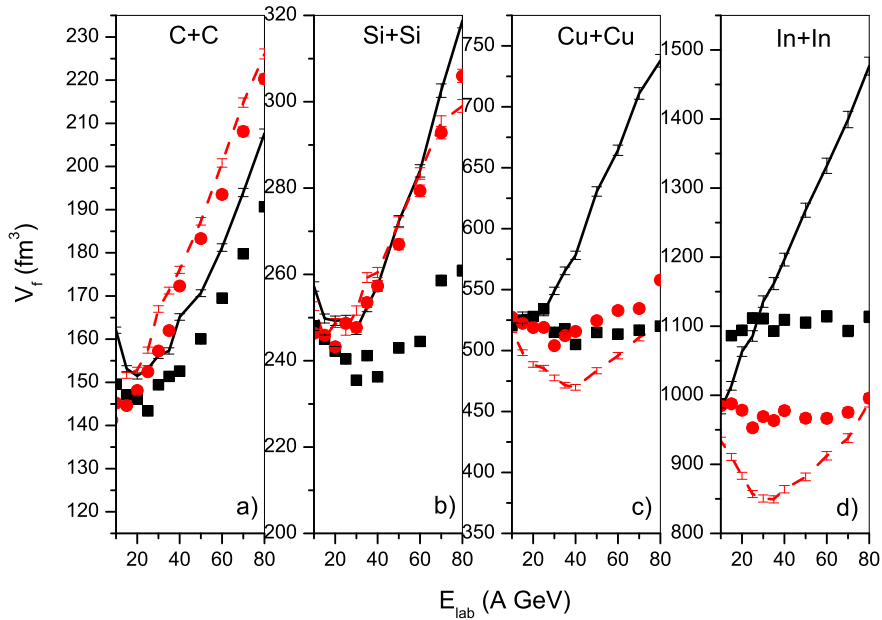}
\caption{Excitation functions of the pion source volume $V_f$ from central C+C (in plot (a)
), Si+Si ( (b) ), Cu+Cu ( (c) ), and In+In ( (d) ) collisions at two $k_T$ bins $0-100$
MeV$/c$ (solid lines and squares) and $100-200$ MeV$/c$ (dashed lines and circles). A
pion-pair rapidity cut $|Y_{\pi\pi}|<0.5$ is employed. Calculations with cascade mode are
shown with symbols while calculations with SM-EoS are shown by the different lines.}
\label{fig4}
\end{figure}

Fig.\ \ref{fig4} shows the excitation function of the pion source
volume $V_f$ ($=(2\pi)^{3/2}R_LR_S^2$) at freeze-out (calculated as
\cite{Adamova:2002ff}) from central C+C (in plot (a) ), Si+Si ( (b)
), Cu+Cu ( (c) ), and In+In ( (d) ) collisions. First of all, for
the calculations in the cascade mode, the excitation function of
$V_f$ shows a minimum only for light systems (especially for the
middle-sized Si+Si system). In the In+In case, the minimum is not
seen in any $k_T$ bins. It implies that the occurrence of the
minimum $V_f$ is system-size dependent if the cascade mode is
adopted in calculations. Secondly, let us explore the results with
the inclusion of mean-field potentials (lines), in C+C case, the
excitation function of $V_f$ shows a minimum at low SPS energies for
the calculation at lower $k_T$ bin (the $V_f$ value decreases about
$7\%$ from the beam energy 10A Gev to 20A GeV). This minimum keeps
for heavier systems Si+Si, Cu+Cu, and In+In, but, the valley within
the lower $k_T$ becomes shallow and disappear while that in the
higher $k_T$ bin starts to become important (the amplitude of the
decrease of $V_f$ from 10A GeV to 35A GeV is about $9\%$).
Therefore, the minimum seems always to be present (although in
different $k_T$ bins) in the excitation function of the pion source
volume $V_f$ of all systems for all calculations with the mean-field
potentials. It implies that the occurrence of the minimum $V_f$ is
$k_T$ dependent but not system-size dependent if the mean-field
potentials are considered in calculations. Finally, the minimum, if
it is present, lies always around the beam energy $30A$ GeV. This is
because the dominant mechanism of the particle production starts to
change from the decay of resonances to the fragmentation of strings
\cite{Weber:1998zb}. Although the transport of free quark degree of
freedom is not considered in current version of the UrQMD, the
mean-field potentials for the pre-formed hadrons may partly
compensate the lack of pressure from the (missing) partonic stage.
The decrease of the pion source volume before the minimum is due to
the fact that more pions starts to be produced from the excitation
and fragmentation of strings which happens at an earlier stage of
the collision, i.e. at a smaller freeze-out size. With the further
increase of the beam energy, the increase of the pion source volume
is due to the increase of pion yields at high SPS energies. However
this effect is washed out in heavier system since both the
production of pions from the decay of resonance and the rescattering
at the late stage of collisions play important roles.

In summary, we have calculated the HBT correlation functions of negatively charged pions for
several systems from the light C+C, middle-sized Si+Si and Cu+Cu, to the heavy In+In system
with the UrQMD transport model and the correlation after-burner CRAB 3.0. We found that the
non-monotonous energy dependence of the pion freeze-out volume might signal a change in the
degrees of freedom from hadronic to partonic matter around $30A$ GeV. The existence of the
minimum was  explained by the combination and competition of the resonance decay with the
string excitation and fragmentation (which we interpreted as a pre-cursor to a QGP). The
results presented here, especially for the middle-sized systems, are in reach of the
NA61/SHINE experiment at the CERN-SPS.

\section*{Acknowledgements}
We would like to thank S. Pratt for providing the CRAB program and
acknowledge support by the computing server C3S2 in Huzhou Teachers
College. The work is supported in part by the key project of the
Ministry of Education of China (No. 209053), the National Natural
Science Foundation of China (Nos. 10905021,10979023, 10979024), the
Zhejiang Provincial Natural Science Foundation of China (No.
Y6090210), and the Qian-Jiang Talents Project of Zhejiang Province
(No. 2010R10102). This work was supported by the Hessian LOEWE
initiative through the Helmholtz International Center for FAIR (HIC
for FAIR).


\end{document}